\documentclass[prl,twocolumn,showpacs,superscriptaddress,amssymb]{revtex4}

\usepackage{psfrag,graphicx}
\usepackage{dcolumn}
\usepackage{bm}
\usepackage{amsfonts,amssymb,amsmath}        

\usepackage{xcolor}

\newcommand{\be}{\begin{equation}}
\newcommand{\ee}{\end{equation}}
\newcommand{\bq}{\begin{eqnarray}}
\newcommand{\eq}{\end{eqnarray}}

\newcommand{\PR}[1]{\ensuremath{\left[#1\right]}}
\newcommand{\PC}[1]{\ensuremath{\left(#1\right)}}
\newcommand{\chav}[1]{\ensuremath{\left\{#1\right\}}}

\newcommand\T{\rule{0pt}{3.0ex}}       
\newcommand\B{\rule[-1.5ex]{0pt}{0pt}} 

\definecolor{kon}{rgb}{.8,.1,0.3}

\begin{document}

\title{Conformal QED in two-dimensional topological insulators}

\author{Nat\'{a}lia Menezes}
 \affiliation{Institute for Theoretical Physics, Center for Extreme Matter and Emergent Phenomena, Utrecht University, Princetonplein 5, 3584 CC Utrecht, the Netherlands}
\author{Giandomenico Palumbo}
 \affiliation{Institute for Theoretical Physics, Center for Extreme Matter and Emergent Phenomena, Utrecht University, Princetonplein 5, 3584 CC Utrecht, the Netherlands}
\author{Cristiane Morais Smith}
 \affiliation{Institute for Theoretical Physics, Center for Extreme Matter and Emergent Phenomena, Utrecht University, Princetonplein 5, 3584 CC Utrecht, the Netherlands}
 \affiliation{Wilczek Quantum Center, Zhejiang University of Technoloogy, Hangzhou 310023, China}

\date{\today}  

\begin{abstract}
{It has been shown recently that local four-fermion interactions on the edges of two-dimensional time-reversal-invariant topological insulators give rise to a new non-Fermi-liquid phase, called helical Luttinger liquid (HLL).
In this work, we provide a first-principle derivation of this non-Fermi-liquid phase based on the gauge-theory approach.
Firstly, we derive a gauge theory for the edge states by simply assuming that the interactions between the Dirac fermions at the edge are mediated by a quantum dynamical electromagnetic field. Here, the massless Dirac fermions are confined to live on the one-dimensional boundary, while the (virtual) photons of the U(1) gauge field are free to propagate in all the three spatial dimensions that represent the physical space where the topological insulator is embedded.
We then determine the effective 1+1-dimensional conformal field theory (CFT) given by the conformal quantum electrodynamics (CQED). By integrating out the gauge field in the corresponding partition function, we show that the CQED gives rise to a 1+1-dimensional Thirring model. The bosonized Thirring Hamiltonian describes exactly a HLL with a parameter K and a renormalized Fermi velocity that depend on the value of the fine-structure constant $\alpha$.}

\end{abstract}

\pacs{71.10.Pm,12.20.-m,71.27.+a,72.25.-b}
\maketitle

\textsl{{\bf Introduction:--}}. Topological insulators represent a large family of materials characterized by gapped bulks and metallic edge states. The topological quantum numbers associated to the bulk depend on the discrete symmetries of the microscopic Hamiltonians, such as time-reversal, particle-hole and chiral symmetries \cite{Altland-Zirnbauer,Ryu1}. Further spatial (crystalline) symmetries have been proposed in order to extend the periodic table of topological free-fermion systems \cite{Juricic, Fu}, and more recently inversion symmetry has also gathered some attention \cite{Bernevig1,Miert}. However, time-reversal-invariant topological insulators are certainly the most studied so far \cite{Hasan,Bernevig-book}. These time-reversal-invariant topological insulators were theoretically proposed to occur in two-dimensional models involving a strong spin-orbit interaction \cite{Kane-Mele, BHZ}, and were then experimentally observed in HgTe quantum wells \cite{Konig}. The spin-orbit interaction locks the spin and the chirality together and produces counter-propagating edge currents, giving rise to the quantum spin Hall effect.
These topologically protected edge modes are right-handed and left-handed Dirac modes that always come in pairs, in agreement with the time-reversal symmetry of the bulk. Their dynamics is consistently described by a 1+1-dimensional massless Dirac theory. 

Moreover, it has been shown that local four-fermion interactions on the edge can transform the free-fermion phase into a new non-Fermi-liquid phase, called helical Luttinger liquid (HLL) \cite{Bernevig2,Moore}.
In this picture, the strength of the interactions is encoded in the Luttinger parameter $K$, which depends on the value of the coupling constant $g$ of the four-fermion term. Although many studies have pointed out for which values of $K$ the interactions are relevant, it is still unclear how the constant $g$ is related to the microscopic properties of the Dirac edge modes, such as their spin, electric charge, etc. The relevant open question is whether there is any fundamental way to derive the HLL from the universal properties of topological insulators.
\begin{figure}[!ht]
	\centering
		\includegraphics[width=0.42\textwidth]{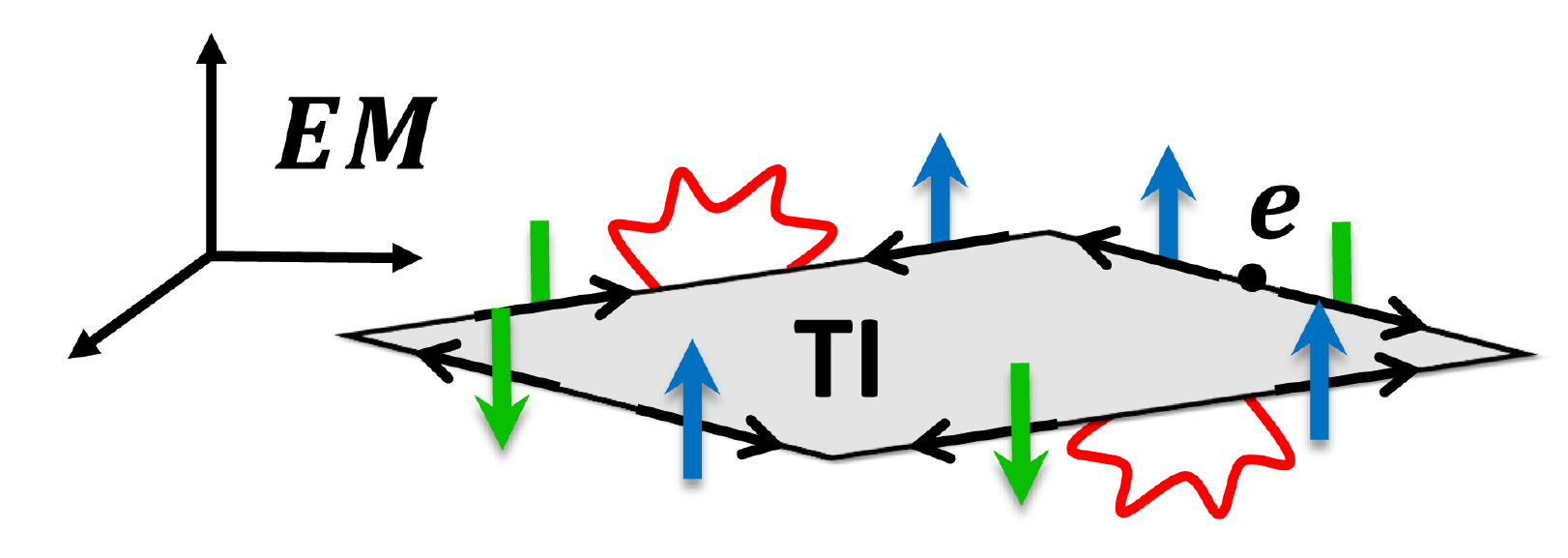}
	\caption{The red wavy lines represent the virtual photons that are free to propagate in all the three spatial dimensions, while the massless Dirac fermions with electric charge $e$ are confined on the one-dimensional boundary of the topological insulator. The arrows at the edges indicate the propagation of the topologically protected right- and left-handed chiral modes.}
	\label{Fig1}
\end{figure}

The main goal of this paper is to provide an answer to this question. Firstly, we derive a gauge theory for the edge states of two-dimensional (2D) time-reversal-invariant topological insulators by simply assuming that the interactions between the Dirac fermions are mediated by a quantum dynamical electromagnetic field. Importantly, the massless Dirac fermions are confined to live on the one-dimensional boundary, while the quantum excitations (i.e. the virtual photons) of the $U(1)$ gauge field are free to propagate in all the three spatial dimensions that represent the physical space where the topological insulator is embedded, see Fig.~\ref{Fig1}. This basic idea allows us to determine the effective 1+1-dimensional gauge theory, which is given by the sum of a conformal quantum electrodynamics (CQED) \cite{Giombi1,Giombi2} and the 1+1-dimensional massless QED, also known as the Schwinger model \cite{Schwinger}. In this work, we focus on the conformal field theory (CFT) because it preserves the dimensionality of both, the electric charge and the gauge field of the projected 3+1-dimensional QED from which the CQED will be derived. This dimensional-reduction method has been already used in studies of graphene \cite{Morais-Smith1,Morais-Smith2} and related 2D massive Dirac systems, such as silicene and transition metal dicalcogenides \cite{Morais-Smith4}, but to the best of our knowledge, it has not yet been employed in the description of one-dimensional systems, such as the edge currents of topological insulators \cite{footnote, Miransky}. Furthermore, by integrating out the gauge field in the corresponding partition function, we find that this gauge theory gives rise to a 1+1-dimensional Thirring model \cite{Thirring}. We then demonstrate that the bosonized version of the interacting-fermion Hamiltonian describes exactly a HLL with a Luttinger parameter $K$ and a renormalized Fermi velocity that depend on the value of the fine-structure constant $\alpha$. Thereby, this work provides a first-principle derivation of a non-Fermi-liquid phase based on the gauge-theory approach.

\textsl{{\bf Conformal QED on the boundary of topological insulators:--}}
We start by considering two-dimensional time-reversal invariant topological insulators in class AII \cite{Ryu1}. They have a gapped bulk and topologically protected Dirac edge modes. These systems realize the quantum spin Hall effect, i.e. the chirality of the Dirac edge modes is locked to the spin, which is preserved due to the time-reversal symmetry. Thus, the dynamics of the edge modes can be described by a 1+1-dimensional massless Dirac theory with a two-component Dirac spinor $\psi=(\psi_{R},\psi_{L})^{T}$, where $\psi_{R}$ and $\psi_{L}$ are the right-handed spin-up and left-handed spin-down chiral modes, respectively. 
It was theoretically proposed in Refs.~\cite{Bernevig2,Moore} and experimentally confirmed in Ref.~\cite{Du} that these topological insulators can support HLLs on the boundary due to the presence of unavoidable electron-electron interactions. These non-Fermi liquid phases fully preserve the time-reversal symmetry and are formally described by the free Dirac theory plus suitable four-fermion interactions. 
We now show that this model and the corresponding HLL can be derived from a gauge theory by simply assuming that the electrically charged propagating Dirac fermions on the edge interact through a quantum dynamical electromagnetic field $A_{\rho}$. The essential point of our approach is that the massless Dirac fermions are confined on the one-dimensional boundary, whereas the quantum excitations (i.e. photons) of the electromagnetic field are free to propagate in all the three spatial dimensions, as shown in Fig.~\ref{Fig1}. The corresponding covariant QED action reads
\begin{align}
S_{{\rm QED}}[A_{\rho},\bar{\psi}, \psi] =i\hbar\int d^{2}r\,\bar{\psi}\gamma^{\mu}\partial_{\mu}\psi \nonumber \\
-\int d^{4}r\, \left(\frac{\varepsilon_{0}c}{4}\,F_{\rho\beta}F^{\rho\beta} + ej_{3+1}^{\rho}A_{\rho}\right), \label{QED4}
\end{align}
where $d^{2}r=v\, dx\, dt$ and $d^{4}r=c\, dx\, dy\, dz\, dt$, with $v$ and $c$ the Fermi velocity and the speed  of light, respectively. $\hbar$ is the Planck constant divided by $2\pi$, $e$ is the electric charge carried by each fermion, $\varepsilon_{0}$ is the vacuum dielectric constant, $\gamma^{\mu}$ are $2\times 2$ Dirac matrices with $\mu=0,1$, $F_{\rho\beta}=\partial_{\rho}A_{\beta}-\partial_{\beta}A_{\rho}$ is the field-strength tensor, $j^{\rho}_{3+1}=\bar{\psi}\gamma^{\rho}\psi$, and $\bar{\psi}=\psi^{\dagger}\gamma^{0}$ with $\rho,\beta=0,1,2,3$.
The effective interaction felt by the massless Dirac fermions due to the gauge field can be obtained by integrating out the $A_{\rho}$-field in the partition function $\mathcal{Z}$, i.e.
\begin{eqnarray}
\mathcal{Z} &=&\int \mathcal{D}\bar{\psi}  \int \mathcal{D}\psi \int \mathcal{D}A_{\rho} \exp \left( i\,S_{\rm QED}\right) \nonumber \\
&=& \int \mathcal{D}\bar{\psi}  \int \mathcal{D}\psi \exp\PC{i\,S_{{\rm eff}}[\bar{\psi},\psi]}, \label{Z}
\end{eqnarray}
where $S_{\rm eff}=S_{D}+S_{\rm int}$ is the effective action, with $S_{D}$ the free Dirac action and $S_{{\rm int}}$ the interaction term, given by
\begin{eqnarray}
S_{\rm int} = -\frac{e^{2}}{2\varepsilon_{0}c}\int d^{4}r d^{4}r' j^{\rho}_{3+1}(r) \frac{1}{(-\Box)} j_{\rho}^{3+1}(r'), \label{Eff}
\end{eqnarray}
where $\Box$ is the d'Alembertian operator.
Now, by imposing the following constraint on the matter current
\begin{equation}
j^{\rho}_{3+1}(t,x,y,z)=j^{\rho}_{1+1}(t,x)\delta(y)\delta(z),\label{constraint}
\end{equation}
we create the dimensional mismatch between the Dirac fermions and the virtual photons, preserving the 3+1 spacetime dimensionality of the electromagnetic field. Hence, by inserting Eq.~(\ref{constraint}) into Eq.~(\ref{Eff}), we get
\begin{eqnarray}
S_{\rm int} &=& -\frac{e^{2}}{2\varepsilon_{0}c}\int d^{2}r d^{2}r' j^{\mu}_{1+1}(r) \PR{\frac{1}{(-\Box)}}_{**} j_{\mu}^{1+1}(r'),\qquad \label{Eff2}
\end{eqnarray}
where the symbol $**$ means that we need to evaluate the Green's function at $y=y'=0$ and $z=z'=0$. To evaluate Eq.~(\ref{Eff2}), we first write the Fourier transform of the Green's function
\begin{align}
\frac{1}{\PC{-\Box}} = -\Box_r \int \frac{d^{4}k}{(2\pi)^{4}} \frac{e^{ik\cdot(r-r')}}{(k^{2})^{2}}, 
\end{align}
where $\Box_r$ is the d'Alembertian in terms of the coordinates. We integrate over the momenta $k$ and then we impose the above constraints on the coordinates. We then find (see Supplemental Material for details)
\begin{eqnarray}
\PR{\frac{1}{\PC{-\Box}}}_{**} &=& \frac{1}{2\pi}\delta(x-x')\delta(t-t')+\frac{1}{(2\pi)^{3/2}}\frac{1}{\Box_{1+1}},\qquad \label{GF}
\end{eqnarray}
where $\delta(x-x')$ and $\delta(t-t')$ are two Dirac delta functions and $\Box_{1+1}$ is the d'Alembertian in 1+1 dimensions. Notice that in Refs.~\cite{Miransky,Teber}, a finite-size regulator for the Dirac delta function in Eq.~(\ref{constraint}) was introduced. This result agrees with ours in the limit when the finite-size regulator is removed.

The replacement of the terms in Eq.~(\ref{GF}) in the effective interaction term (\ref{Eff2}) leads to 
\begin{eqnarray}
S_{\rm int} = -\frac{e^{2}}{4\pi\varepsilon_{0}c}\int dt dx j^{\mu}_{1+1}(x,t)j_{\mu}^{1+1}(x,t) \nonumber \\ -\frac{e^{2}}{(2\pi)^{3/2}\varepsilon_{0}c}\int dt dt' dx dx' j^{\mu}_{1+1}(x,t)\frac{1}{\Box}j_{\mu}^{1+1}(x',t'). \label{Eff3}
\end{eqnarray}
An alternative way to obtain this effective interaction is to introduce two independent gauge fields $A^{a}_\mu$ with $a=1,2$, i.e.
\begin{eqnarray}
S_{{\rm CQED+Maxwell}}[A^{a}_{\mu},\bar{\psi}, \psi] = \hspace{2.0cm}\nonumber \\ \int d^{2}r\, \left(i\hbar\bar{\psi}\gamma^{\mu}\partial_{\mu}\psi - e j_{1+1}^{\mu}A^{1}_{\mu}-\frac{\pi\varepsilon_{0}c}{2} F^{1}_{\mu\nu}\frac{1}{\Box}F_{1}^{\mu\nu}\right.\nonumber \\ \left.-\bar{e} j_{1+1}^{\mu}A^{2}_{\mu}-\frac{\pi\varepsilon_{0}c\sqrt{2\pi}}{4}\,F^{2}_{\mu\nu}F_{2}^{\mu\nu}\right), \label{QED1-1}
\end{eqnarray}
which replaces the action (\ref{QED4}) and represents the main result of this work. Integrating out the $A^{a}_{\mu}$-fields in Eq.~(\ref{QED1-1}), one obtains exactly Eq.~(\ref{Eff3}). The pseudo-differential operator in the kinetic term of the $A^{1}_{\mu}$-field adjusts its dimensionality, such that the coupling constant $e$ remains dimensionless, while $\bar{e}$ is a dimensionful bare constant. 

Furthermore, from our result (\ref{QED1-1}) we can derive two well-known exactly solvable models in 1+1-dimensions: by integrating out the $A^{1}_{\mu}$-field, we obtain the Thirring model \cite{Thirring}, whereas the Lagrangian for the $A^{2}_{\mu}$-field can be identified with the Schwinger model \cite{Schwinger}. Because the photon does not have dynamical degrees of freedom in 1+1-dimensional QED \cite{Ruiz-Estrada}, from now on we neglect the $A^{2}_{\mu}$-field and focus only on the conformal sector of Eq.~(\ref{QED1-1}), which generates the Thirring model. 

Now, we stress some relevant properties of this 1+1-dimensional CQED, by showing its connection with the 2+1-dimensional PQED derived in Ref.~\cite{Marino} through the dimensional-reduction procedure. First of all, it is a fully dynamical theory, i.e. both fermions and photons propagate differently from the Maxwell theory, which is intrinsically topological in 1+1 dimensions. Moreover, in the latter, the electric charge $e$ is a dimensionful parameter, while in the CQED it is dimensionless, like the $e$ defined in the 3+1-dimensional QED. The fact that the coupling constant remains dimensionless makes perturbative studies more reliable. 
In 2+1 dimensions, the PQED (see the Table~I) has similar properties, being unitary \cite{Morais-Smith3}, dynamical and renormalizable like (1+1)-dimensional CQED and (3+1)-dimensional QED. Importantly, both PQED and CQED can be derived from the standard QED through the dimensional-reduction approach developed in Ref.~\cite{Marino} and also employed in this work. This is the reason why all these three theories share several properties, even though they are defined in different spacetime dimensions. 

\begin{table}[!ht]
\begin{tabular}{ c|c}
U(1) gauge theories\ &\ Bosonic Lagrangians  \\
\cline{1-2}
1+1 CQED & $-\frac{\pi}{2} F_{\mu\nu}\frac{1}{\Box}F^{\mu\nu}$\T\B  \\
2+1 PQED & $-\frac{1}{2}F_{\mu\nu}\frac{1}{\sqrt{\Box}}F^{\mu\nu}$\T\B  \\
3+1 QED & $-\frac{1}{4}F_{\mu\nu}F^{\mu\nu}$\T\B  \\
\end{tabular}
\caption{The bosonic sector of the QED, PQED and CQED in the second column for $\varepsilon_{0}=c=1$. In lower dimensions, the Maxwell theory is replaced by suitable versions that contains pseudo-operators, i.e. $(\partial^{2})^{-\eta}$ with $\eta=1\ {\rm or}\ 1/2$, to adjust and preserve the dimensionality of the coupling constant $[e]=1$. This means that QED, PQED and CQED are renormalizable theories.}
\end{table}

\textsl{{\bf Thirring model and helical Luttinger liquid:--}} Here, we derive in a straightforward way the HLL from our effective field-theory model. The fermionic kinematical term in Eq.~(\ref{QED4}), together with the local interaction term in Eq.~(\ref{Eff3}), allow us to write the purely effective fermionic action
\begin{equation}
S^{{\rm eff}}_{1+1} =\int d^{2}r\,\left[ i\hbar\bar{\psi}\gamma^{\mu}\partial_{\mu}\psi - g(\bar{\psi}\gamma^{\mu}\psi)^{2}\right], \label{FEL}
\end{equation}
which can be recognized as the massless Thirring model \cite{Thirring}, with the coupling constant $g=e^{2}/4\pi\varepsilon_{0}c$. 
The corresponding Hamiltonian is then calculated by employing a Legendre transformation,
\begin{eqnarray}
H_{\rm eff}= v\int dx \left[ i\hbar\PC{\psi_{R}^{\dagger}\partial_{x}\psi_{R}-\psi_{L}^{\dagger}\partial_{x}\psi_{L}}\right. \nonumber \\
\left.+ \frac{e^{2}}{\pi\varepsilon_{0}c} \psi^{\dagger}_{R}\psi_{R}\psi^{\dagger}_{L}\psi_{L} \right], \label{Hamiltonian} 
\end{eqnarray}
where the interaction term is nothing but the forward scattering, and we have used the chiral basis with $\psi=(\psi_{R},\psi_{L})^{T}$, with the fermion operators satisfying usual anti-commutation relations.
The bosonization of Eq.~(\ref{Hamiltonian}) is straightforward, and we obtain
\begin{eqnarray}
H_{\rm eff}^{\rm bos}&=& \tilde{v} \int dx \PR{\frac{1}{K} \PC{\partial_x\varphi}^{2} +  K\PC{\partial_x\theta}^{2}}, \label{FullHB}
\end{eqnarray}
which is nothing but the Hamiltonian of the HLL, with the scalar fields $\varphi=(\phi_{R}+\phi_{L})/\sqrt{2}$ and $\theta=(\phi_{R}-\phi_{L})/\sqrt{2}$. Here, the bosonization rules read
\begin{eqnarray}
\psi_R=\frac{1}{\sqrt{2\pi}}\,e^{-i\sqrt{4\pi}\phi_R}, \hspace{0.5cm}\psi_L=\frac{1}{\sqrt{2\pi}}\,e^{i\sqrt{4\pi}\phi_L},
\end{eqnarray}
with the Luttinger parameter $K$ and the renormalized velocity $\tilde{v}$ respectively given by
\begin{eqnarray}
K&=& \sqrt{\PC{1-\frac{2\alpha}{\pi}}\PC{1+\frac{2\alpha}{\pi}}^{-1}}, \label{Kparameter} \\
\tilde{v} &=& \hbar v\sqrt{1-\frac{4\alpha^{2}}{\pi^{2}}}, \label{velocity}
\end{eqnarray}
where $\alpha \equiv e^{2}/4\pi\hbar\varepsilon v$ is a measure of the strength of the electron-electron interaction, also known as the fine-structure constant. Because $\alpha$ is an observable that depends on the material, i.e. on the dielectric constant of the medium, we replaced $\varepsilon_{0} \rightarrow \varepsilon$ and $v$ is the velocity of the fermions when they propagate in this material. 
Thus, due to gauge principle and to the projection from QED to CQED, we have been able to derive the HLL on the boundary of the topological insulator. Moreover, we have determined the value of the Luttinger parameter and the renormalized velocity, which depend, in our framework, only on the generic properties of the Dirac modes, i.e. the value of their electric charge, the Fermi velocity and the dielectric constant by means of the fine-structure constant $\alpha$.

\textsl{{\bf Luttinger-parameter discussion:--}}
The parameter $K$ in the HLL defines different regimes of the interaction, which changes from repulsive ($K<1$), passing through non-interacting ($K=1$), to attractive ($K>1$) \cite{Giamarchi}. Nonetheless, how this parameter relates to fundamental properties of the materials was still unclear. In Refs.~\cite{Kane2009, Maciejko}, a formula that connects $K$ with $\alpha$ is derived by employing perturbation theory with either the Kondo or the backscattering interaction.
\begin{figure}[t]
\begin{center}
\includegraphics[scale=0.41]{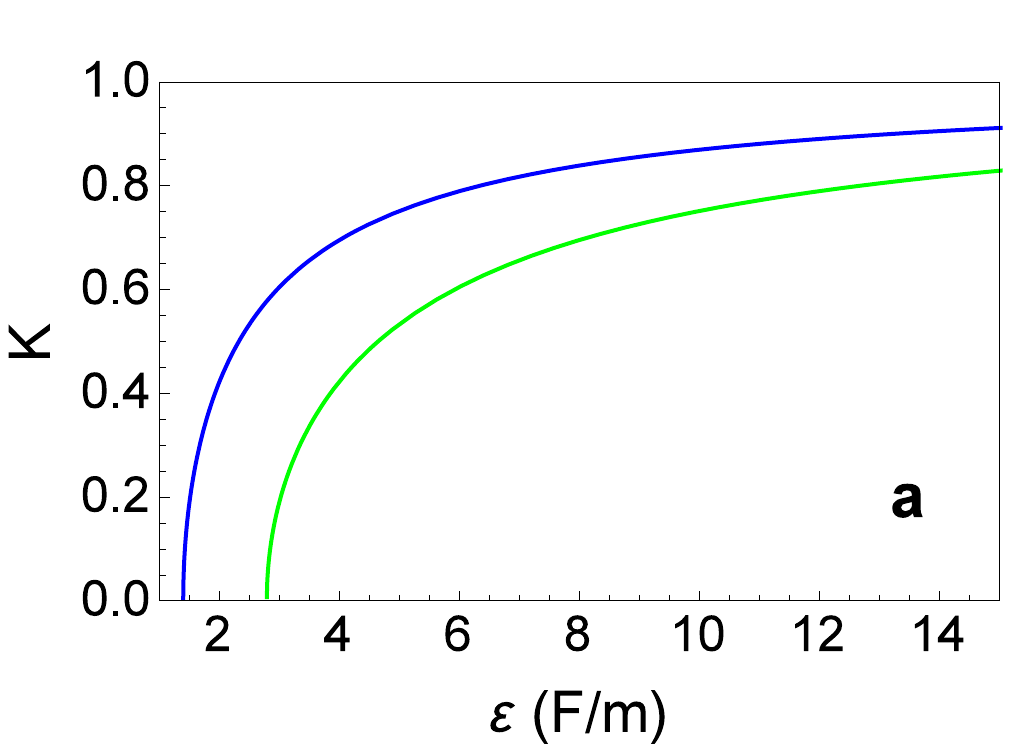}
\includegraphics[scale=0.41]{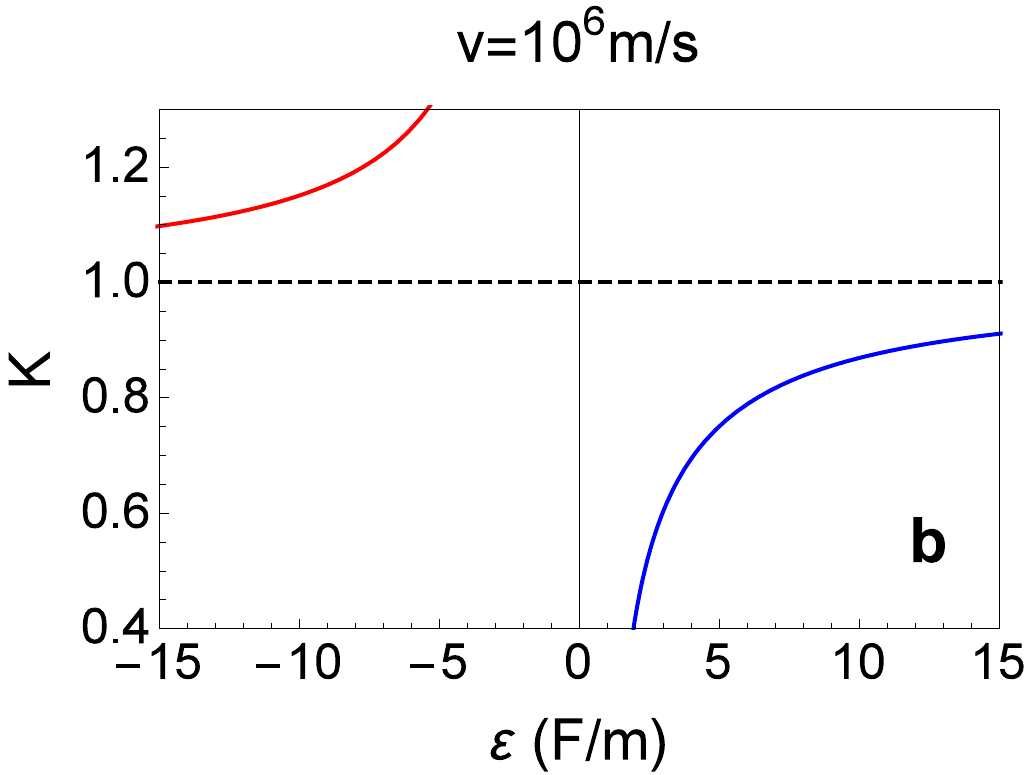}
\end{center}
\caption{(Color online) Luttinger parameter $K$ dependence on the dielectric constant $\varepsilon$ for fixed values of the Fermi velocity $v$. (a) The blue (black) and green (grey) curves are for $v=10^{6}$m/s and $v=5\times 10^{5}$m/s, respectively, and they indicate that for sufficiently large values of $\varepsilon$, the system becomes non-interacting ($K=1$), while for smaller values of $\varepsilon$ the interaction is repulsive ($K<1$). (b) A proposal to obtain attractive interactions $K>1$ by changing the sign of the dielectric constant (red/grey curve) for a sample with $v=10^{6}$m/s.}\label{Kdependence}
\end{figure}
In Sec.~IV, we have presented an \textit{exact} derivation of the Luttinger parameter, which is found to depend on the strength of the electron-electron interaction $\alpha$. 

Now, we compare our results with a prior theoretical prediction proposed in Refs.~\cite{Kane2009,Maciejko}, $K=\PR{1+(8\alpha/\pi)\ln(d/\ell)}^{-1/2}$. Here, $d$ is the distance from the quantum wells to a closeby metallic gate, and $\ell$ acts as a cutoff for short distances. This dependence of the parameter $K$ on $\alpha$ was obtained at the level of perturbation theory on the HLL Hamiltonian, i.e. additional interaction terms had to be taken into account, such as the Kondo or the backscattering interaction. Using the values of the parameters reported experimentally for HgTe quantum wells, $v\approx 5.5\times 10^{5}$ m/s \cite{Konig,Gusev}, $\varepsilon=15$ F/m \cite{Kane2009,Hankiewicz}, $d=150$ nm and $\ell={\rm max} \{30,12\}$ nm \cite{Du}, they find $K\approx0.8$ \cite{Kane2009}. Within our model, which depends only on $\alpha$, we obtain $K\approx 0.84$.

Notice that our approach does not involve the backscattering term, which induces further corrections to the parameter $K$, as seen in the case of InAs/GaSb quantum wells \cite{Du}. This implies that our theoretical prediction applies to materials that have weak backscattering and high Fermi velocities, such as HgTe \cite{Du2}. Nevertheless, the backscattering term can be obtained within our approach upon considering the \textit{massive} Thirring model (see Supplemental Material). Other possible 2D topological insulators that would be good candidates to test our theoretical proposal are plumbene monolayers \cite{Material1} and germanene films \cite{Material2}. The Fermi velocity in these materials has the same order of magnitude as that in HgTe, indicating that backscattering might not be so relevant.

Furthermore, we show how to tune $K$ in order to obtain different regimes of interaction. From Eq.~(\ref{Kparameter}), we notice that to change $K$ we can either change $v$ or the dielectric constant of the medium. In Fig.~\ref{Kdependence}~\textbf{a}, we depict the dependence of $K$ on the dielectric constant $\varepsilon$ in the range [1-15] {\rm F/m}, for a fixed velocity $v=10^{6}$m/s. In the asymptotic limit where $\varepsilon \rightarrow \infty$ (meaning that we are considering very large values of the dielectric constant, not a mathematical infinity), it would be possible to reach the value of $K=1$. For smaller velocities $v$, the minimum value of the dielectric constant for which $K$ becomes real increases, i.e., for $v=5\times 10^{5}$m/s, e.g., $\varepsilon\approx 2.7-\infty\ {\rm F/m}$. On the other hand, if we consider negative values of the dielectric constant by placing the topological insulator on top of a meta-material, then it is possible to switch from repulsive to attractive interactions, i.e., $K(x) \rightarrow K(-|x|) = \sqrt{(1+|x|)/(1-|x|)}$ with $x=2\alpha/\pi$. We illustrate this situation in Fig.~\ref{Kdependence}~\textbf{b}. The dielectric constant of the medium here plays the same role of Feshback resonances in ultracold atoms, which allow to tune the interaction parameter from the repulsive to the attractive regime \cite{Feshback}. 

\textsl{{\bf Conclusions:--}} In this paper, we derived a gauge theory on the boundary of two-dimensional time-reversal-invariant topological insulators. Our starting point was to assume that the interactions between the charged one-dimensional Dirac fermions at the edge are mediated by a quantum dynamical electromagnetic field, where the virtual photons are free to propagate in all the three spatial dimensions. By implementing a dimensional-reduction procedure, we derived the corresponding CQED, which describes the HLL. Thus, we provided a first-principle description of this non-Fermi-liquid phase based on the gauge-theory approach, in which the parameter $K$ depends on the fine-structure constant. We emphasize that our approach is \textit{exact}, i.e. non-perturbative, and has a more vast applicability in condensed-matter physics. In fact, the one-dimensional effective theory derived here also works in the case of nanowires, in which the HLL phase can be easily obtained \cite{Oppen, Egger}, as done for topological insulators.

Our work provides, to the best of our knowledge, the first microscopic derivation of the Thirring model and opens the path to the manipulation of the Luttinger parameter $K$ by modifying the dielectric constant of the substrate on which the one-dimensional system might be deposited. Interestingly, we find that upon the use of a meta-material as a substrate, it is possible to change the interactions from repulsive into attractive. These results might have profound implications for transport properties in nanostructures in particular, and nanotechnology in general. 

\textsl{{\bf Acknowledgments:--}}
This work was supported by CNPq (Brazil) through the Brazilian government project Science Without Borders. The work of G.P. and also C.M.S. is part of the DITP consortium, a program of the Netherlands Organisation for Scientific Research (NWO) that is funded by the Dutch Ministry of Education, Culture and Science (OCW). We are grateful to Eduardo C. Marino, Vladimir Gritsev, Dirk Schuricht, George Japaridze and Lars Fritz for inspiring discussions. \\

\onecolumngrid
\newpage

\section*{\LARGE{Supplemental Material}}
\onecolumngrid
\maketitle

\section{Details of the calculation on the projection from QED in (3+1)D to (1+1)D}

Here, we show the detailed calculation starting from Eq.~(5) to obtain Eq.~(6) in the main text. The Fourier transform of the photon propagator reads
\begin{eqnarray}
\frac{1}{\PC{-\Box}} =  \int \frac{d^{4}k}{(2\pi)^{4}} \frac{e^{ik\cdot(r-r')}}{k^{2}}, \label{1}
\end{eqnarray}
where $k^{2}=k_{x}^{2}+k_{y}^{2}+k_{z}^{2}+\omega^{2}$.
First, we apply the constraint only on the $z$-component ($z=z'=0$) and integrate Eq.~\eqref{1} over $k_{z}$ to obtain
\begin{eqnarray}
\PR{\frac{1}{\PC{-\Box}}}_{*} = \frac{1}{2} \int \frac{d^{3}k}{(2\pi)^{3}}\frac{e^{ik(r-r')}}{\sqrt{k^{2}}}, \label{2}
\end{eqnarray}
which is the known result of PQED [1]. The symbol $*$ means that we already imposed one of the constraints in the interaction term. Now, if one tries to follow the same steps and integrates over $k_{y}$, after applying the constraints on the $y$-component ($y=y'=0$), the integral does not converge unless a cutoff is introduced. However, since our goal is to derive a conformal theory, we do not intend to introduce a new scaling in the theory by means of a cutoff.

We present an alternative way to solve this problem by rewriting Eq.~\eqref{2} as
\begin{eqnarray}
\PR{\frac{1}{\PC{-\Box}}}_{*} = -\frac{\Box_r}{2} \int \frac{d^{3}k}{(2\pi)^{3}}\frac{e^{ik(r-r')}}{\PC{k^{2}}^{3/2}} = -\frac{\Box_r}{4\pi} \int \frac{d^{2}k}{(2\pi)^{2}}e^{ik_{x}(x-x')+i\omega(t-t')}\int_{-\infty}^{\infty} dk_{y}\frac{e^{ik_{y}(y-y')}}{\PC{ k_{x}^{2}+k_{y}^{2}+\omega^{2}}^{3/2}}, \label{3}
\end{eqnarray}
where $\Box_r$ is a differential operator that only applies to the $r$-component. The exponential in $k_{y}$ can be expanded as
\begin{eqnarray}
e^{ik_{y}(y-y')}= \sum_{n=0}^{\infty} \frac{i^{n}k_{y}^{n}(y-y')^{n}}{n!}=1+\sum_{n=1}^{\infty} \frac{i^{n}k_{y}^{n}(y-y')^{n}}{n!}. \label{4}
\end{eqnarray}
We split the contributions for $n=0$ and $n>0$ in the summation of Eq.~\eqref{4} to show explicitly how the contact interaction emerges.

Replacing Eq.~\eqref{4} into Eq.~\eqref{3} and focusing on the integral over $k_{y}$, we find
\begin{align}
\int_{-\infty}^{\infty} dk_{y}\frac{1}{\PC{ k_{x}^{2}+k_{y}^{2}+\omega^{2}}^{3/2}} \PC{1+\sum_{n=1}^{\infty} \frac{i^{n}k_{y}^{n}(y-y')^{n}}{n!}} = \frac{2}{k_{x}^{2}+\omega^{2}} + \sum_{n=1}^{\infty}  \frac{i^{n}(y-y')^{n}}{n!} \frac{[1+(-1)^{n}]\Gamma\PC{1-\frac{n}{2}}\Gamma\PC{\frac{n+1}{2}}}{\sqrt{\pi}\PC{k_{x}^{2}+\omega^{2}}^{1-\frac{n}{2}}}, \label{5}
\end{align}
where the sum is only valid for even values of $n$, and for $n=2$ the Gamma function has a pole. Fortunately, we show later that this pole cancels when one integrates further.  

By replacing Eq.~\eqref{5} into Eq.~\eqref{3}, we find
\begin{eqnarray}
\PR{\frac{1}{\PC{-\Box}}}_{*} = -\frac{\Box_r}{4\pi} \int \frac{d^{2}k}{(2\pi)^{2}}e^{ik_{x}(x-x')+i\omega(t-t')} \chav{\frac{2}{k_{x}^{2}+\omega^{2}} + \sum_{n=1}^{\infty}  \frac{i^{n}(y-y')^{n}}{n!} \frac{[1+(-1)^{n}]\Gamma\PC{1-\frac{n}{2}}\Gamma\PC{\frac{n+1}{2}}}{\sqrt{\pi}\PC{k_{x}^{2}+\omega^{2}}^{1-\frac{n}{2}}}}, \label{6}
\end{eqnarray}
and now we can apply the derivatives to the remaining functions. The first term of Eq.~\eqref{6} generates the local interaction, i.e.
\begin{eqnarray}
-\frac{\Box_r}{2\pi} \int \frac{d\omega }{2\pi}  \int \frac{dk_{x}}{2\pi} \frac{e^{ik_{x}(x-x')+i\omega(t-t')}}{k_{x}^{2}+\omega^{2}}= \frac{1}{2\pi} \delta(x-x')\delta(t-t'), \label{7}
\end{eqnarray}
which appears due to the first contribution of the expansion of Eq.~\eqref{4}. The result obtained in Eq.~\eqref{7} does not depend on whether we consider or not the constraint on the $y$-component. However, this is not the case for the second term of Eq.~\eqref{6}, which gives 
\begin{eqnarray}
&&-\frac{1}{4\pi}\sum_{n\ {\rm even}}^{\infty} \frac{i^{n}\Gamma\PC{1-\frac{n}{2}}\Gamma\PC{\frac{n+1}{2}}}{\sqrt{\pi}(n!)} \Box_r \PR{(y-y')^{n} \int \frac{d^{2}k}{(2\pi)^{2}}\frac{e^{ik_{x}(x-x')+i\omega(t-t')}}{\PC{k_{x}^{2}+\omega^{2}}^{1-\frac{n}{2}}}} \nonumber \\
 = &-&\frac{1}{4\pi}\sum_{n\ {\rm even}}^{\infty} \frac{i^{n}\Gamma\PC{1-\frac{n}{2}}\Gamma\PC{\frac{n+1}{2}}}{\sqrt{\pi}(n!)} \PR{n(n-1)(y-y')^{n-2} \int \frac{d^{2}k}{(2\pi)^{2}}\frac{e^{ik_{x}(x-x')+i\omega(t-t')} }{\PC{k_{x}^{2}+\omega^{2}}^{1-\frac{n}{2}}}} \nonumber \\
&-&\frac{1}{4\pi}\sum_{n\ {\rm even}}^{\infty} \frac{i^{n}(-n/2)\Gamma\PC{-\frac{n}{2}}\Gamma\PC{\frac{n+1}{2}}}{\sqrt{\pi}(n!)} \PR{(y-y')^{n-2} \int \frac{d^{2}k}{(2\pi)^{2}}\frac{e^{ik_{x}(x-x')+i\omega(t-t')}}{\PC{k_{x}^{2}+\omega^{2}}^{-\frac{n}{2}}}}.\label{8}
\end{eqnarray}

Integrating over $\omega$ for both terms in Eq.~\eqref{8}, we find
\begin{eqnarray}
&&\frac{1}{(2\pi)^{2}} \int_{-\infty}^{\infty} dk_{x} e^{ik_{x}(x-x')} \int_{-\infty}^{\infty} d\omega \frac{e^{i\omega(t-t')}}{\PC{k_{x}^{2}+\omega^{2}}^{1-\frac{n}{2}}} \nonumber \\
&&= \frac{2^{(1+n)/2}\sqrt{\pi}}{(2\pi)^{2}\Gamma\PC{1-\frac{n}{2}}} \int_{-\infty}^{\infty} dk_{x} e^{ik_{x}(x-x')} |t-t'|^{(1-n)/2}(k_{x}^{2})^{\frac{n-1}{4}}K_{\frac{n-1}{2}} \PC{|t-t'|\sqrt{k_{x}^{2}}},  \label{9}
\end{eqnarray}
and 
\begin{eqnarray}
&&\frac{1}{(2\pi)^{2}} \int_{-\infty}^{\infty} dk_{x} e^{ik_{x}(x-x')} \int_{-\infty}^{\infty} d\omega \frac{e^{i\omega(t-t')}}{\PC{k_{x}^{2}+\omega^{2}}^{-\frac{n}{2}}} \nonumber \\
&&= \frac{2^{(3+n)/2}\sqrt{\pi}}{(2\pi)^{2}\Gamma\PC{-\frac{n}{2}}} \int_{-\infty}^{\infty} dk_{x} e^{ik_{x}(x-x')} |t-t'|^{-(1+n)/2}(k_{x}^{2})^{\frac{n+3}{4}} K_{\frac{n+1}{2}} \PC{|t-t'|\sqrt{k_{x}^{2}}},  \label{10}
\end{eqnarray}
where $K_{\eta}$'s are modified Bessel functions of the second kind. 
Plugging the results of Eqs.~\eqref{9} and \eqref{10} into Eq.~\eqref{8}, we see that the poles disappear. Moreover, by imposing the constraint on the $y$-component ($y=y'=0$), we observe that all the $n$-even contributions vanish, except $n=2$. For $n=2$, Eq.~\eqref{8} becomes
\begin{eqnarray}
\frac{2^{1/2}\Gamma\PC{\frac{3}{2}}}{4\pi^2}   \int_{-\infty}^{\infty} dk_{x} e^{ik_{x}(x-x')} \frac{ e^{-|t-t'|\sqrt{k_{x}^{2}}}}{|t-t'|} = \frac{1}{(2\pi)^{3/2}} \frac{ 1}{|t-t'|^{2}+|x-x'|^{2}} . \label{11}
\end{eqnarray}

Hence, summing the  results of Eqs.~\eqref{7} and \eqref{11}, we have
\begin{eqnarray}
\PR{\frac{1}{\PC{-\Box}}}_{**} = \frac{1}{2\pi} \delta(x-x')\delta(t-t') + \frac{1}{(2\pi)^{3/2}} \frac{ 1}{(t-t')^{2}+(x-x')^{2}},\label{12}
\end{eqnarray}
where the symbol $**$ means that we took both the $y-$ and the $z-$coordinate constraints into account. Interestingly, the Fourier transform of the second term in Eq.~\eqref{12} is actually
\begin{eqnarray}
\frac{1}{(t-t')^{2}+(x-x')^{2}} = \int \frac{dk_{x}}{2\pi}  \int \frac{d\omega }{2\pi} \frac{e^{ik_{x}(x-x')+i\omega(t-t')}}{\omega^{2}+k_{x}^{2}} \equiv \frac{1}{\Box_{1+1}},\label{step12}
\end{eqnarray}
which then yields an effective interaction composed of a sum of a local and a non-local term, i.e.
\begin{eqnarray}
\PR{\frac{1}{\PC{-\Box}}}_{**} = \frac{1}{2\pi} \delta(x-x')\delta(t-t') + \frac{1}{(2\pi)^{3/2}}\frac{1}{\Box_{1+1}}.\label{13}
\end{eqnarray}

\section{Effective action and 1+1-dimensional Lagrangian}

From the result found in Eq.~\eqref{13}, the effective action reads
\begin{eqnarray}
S_{{\rm eff}} &=& -\frac{e^{2}}{2\varepsilon_{0}c}\int d^{4}r d^{4}r' j^{\mu}_{3+1}(r) \frac{1}{(-\Box)} j_{\mu}^{3+1}(r') \nonumber \\
&=& -\frac{e^{2}}{2\varepsilon_{0}c}\int dt dt' dx dx' j^{\mu}_{1+1}(x,t)\PR{\frac{1}{\PC{-\Box}}}_{**} j_{\mu}^{1+1}(x',t')\nonumber \\
&=& -\frac{e^{2}}{4\pi\varepsilon_{0}c}\int dt dx j^{\mu}_{1+1}(x,t)j^{\mu}_{1+1}(x,t) -\frac{e^{2}}{(2\pi)^{3/2}\varepsilon_{0}c}\int dt dt' dx dx' j^{\mu}_{1+1}(x,t)\frac{1}{\Box}j_{\mu}^{1+1}(x',t'). \label{Seff}
\end{eqnarray}

The two terms in the effective action \eqref{Seff} can be viewed as if the fermions were mediated by two distinct gauge fields in 1+1-dimensions, i.e.
\begin{eqnarray}
\mathcal{L}_{1+1} = i\hbar\bar{\psi}\gamma^{\mu}\partial_{\mu}\psi  - ej^{\mu}A^{1}_{\mu}-\bar{e}j^{\mu}A^{2}_{\mu}-g_{1}F^{1}_{\mu\nu}\frac{1}{\Box}F_{1}^{\mu\nu} -g_{2}F^{2}_{\mu\nu}F_{2}^{\mu\nu}, \label{Lagrangian}
\end{eqnarray}
where $g_{1}=\pi\varepsilon_{0}c/2$ and $g_{2}=\pi\varepsilon_{0}c\sqrt{2\pi}/4$ are dimensionless constants. By integrating out the $A^{1}_{\mu}$-field we obtain the Thirring model [2], whereas the Lagrangian for $A^{2}_{\mu}$ gives us the Schwinger model [3]. Both models are exactly solvable in 1+1-dimensions.
Notice that $\bar{e}$ is a dimensioful bare constant, which is in agreement with the Schwinger model.

The correspondence between Eqs.~\eqref{Seff} and \eqref{Lagrangian} can be seen explicitly by squaring the gauge fields $A^{1}_{\mu}$ and $A^{2}_{\mu}$. In this manner, we obtain the following effective interactions between the matter currents
\begin{eqnarray}
g_{1}\PC{ -F^{1}_{\mu\nu}\frac{1}{\Box}F_{1}^{\mu\nu} - \frac{ej^{\mu}A^{1}_{\mu}}{g_{1}} } =  2g_{1}\PC{ A^{1}_{\mu}A_{1}^{\mu} - \frac{ej^{\mu}A^{1}_{\mu}}{2g_{1}} }   = 2g_{1}\PC{ A^{1}_{\mu}A_{1}^{\mu} - \frac{2ej^{\mu}A^{1}_{\mu}}{4g_{1}} + \frac{e^{2}j^{\mu}j_{\mu}}{16g_{1}^{2}} } - \frac{e^{2}j^{\mu}j_{\mu}}{8g_{1}}, \nonumber
\end{eqnarray}
 and 
\begin{eqnarray}
g_{2}\PC{ -F^{2}_{\mu\nu}F_{2}^{\mu\nu} - \frac{\bar{e}j^{\mu}A^{2}_{\mu}}{g_{2}} } =  2g_{2}\PC{ A^{2}_{\mu}\partial^{\nu}\partial_{\nu}A_{2}^{\mu} - \frac{\bar{e}j^{\mu}A^{2}_{\mu}}{2g_{2}} }  = 2g_{2}\PC{ A^{2}_{\mu}\partial^{2}A_{2}^{\mu} - \frac{2\bar{e}j^{\mu}A^{2}_{\mu}}{4g_{2}} + \frac{\bar{e}^{2}j^{\mu}\partial^{-2}j_{\mu}}{16g_{2}^{2}} } - \frac{\bar{e}^{2}}{8g_{2}}j^{\mu}\frac{1}{\Box}j_{\mu}, \nonumber
\end{eqnarray}
which for $g_{1}=\pi\varepsilon_{0}c/2$ and $g_{2}=\pi\varepsilon_{0}c\sqrt{2\pi}/4$ reproduce those two terms in Eq.~\eqref{Seff}. 

\section{Masses in the Thirring model and the backscattering interaction}

To investigate the properties of the edge states in presence of an external Zeeman field, which breaks time-reversal symmetry, one may add a mass (i.e. $m \bar{\psi}\psi$) to the Dirac fermions in Eq.~(1) of the main text. The parameter $m$ is proportional to the intensity of the Zeeman field, which we consider for simplicity constant in modulus and direction. Because this mass term is not affected by the dimensional-reduction procedure, it also appears in the Thirring model, generating a gap in the boundary modes. In the Hamiltonian picture, this massive term is written as
\begin{eqnarray}
H_m = m \int dx\PC{\psi^{\dagger}_{R}\psi_{L}+\psi^{\dagger}_{L}\psi_{R}}.\label{Fmass}
\end{eqnarray}
Now, by using the bosonization rules with the Klein factors properly defined [4], the above massive term becomes
\begin{eqnarray}\label{cosine}
H_{m}^{\rm bos}=\frac{m }{\pi}\int dx \cos\PC{\sqrt{8\pi}\varphi}. \label{Hm}
\end{eqnarray}
For the bosonic representation, this cosine term, when localized in a small region, acts as a boundary in the system, changing the fermionic orientation of propagation. The existence of such contribution leads to the study of the renormalization group of the Sine-Gordon model, as already analyzed in Ref.~[5]. Note that the above term looks similar to the one obtained in Ref.~[6], induced by the Umklapp scattering.

At a theoretical level, another possible massive term is $i\Delta\bar{\psi}\gamma^{5}\psi$ with $\gamma^{5}=\gamma^{0}\gamma^{1}$. The coefficient $\Delta$ is known as the chiral mass and it adds to the fermionic Hamiltonian the following contribution 
\begin{eqnarray}
H_{\Delta} = i\Delta \int dx \PC{\psi^{\dagger}_{R}\psi_{L}-\psi^{\dagger}_{L}\psi_{R}},\label{Chiralmass}
\end{eqnarray}
which mainly differs from the usual Dirac mass by a factor minus between $\psi^{\dagger}_{R}\psi_{L}$ and its conjugated. The minus sign in Eq.~\eqref{Chiralmass} leads to a bosonized Hamiltonian containing an interaction term that depends on a sine function instead of a cosine, i.e.,  
\begin{eqnarray}
H_{\Delta}^{\rm bos}= - \frac{\Delta}{\pi} \int dx \sin\PC{\sqrt{8\pi}\varphi}. \label{HDelta}
\end{eqnarray}
Here, we can easily recover the standard potential in Eq.~\eqref{cosine} after a constant shift of the scalar field, i.e. $\varphi\rightarrow \varphi - \pi/2$.

\section*{References}
 
[1] E. C. Marino, Nucl. Phys. B \textbf{408}, 551 (1993).

[2] W. Thirring, Ann. Phys. \textbf{3}, 91 (1958).

[3] J. S. Schwinger, Phys. Rev. \textbf{128}, 2425 (1962).
 
[4] T. Lee, J. Korean Phys. Soc. \textbf{68}, 1272 (2016).

[5] J. Maciejko, C. Liu, Y. Oreg, X.-L. Qi, C. Wu and S.-C. Zhang, Phys. Rev. Lett. \textbf{102}, 256803 (2009).

[6] C. Wu, B. A. Bernevig and S.-C. Zhang, Phys. Rev. Lett. \textbf{96}, 106401 (2006).


\begin{thebibliography}{100}
\bibitem{footnote} In Ref.~\cite{Miransky}, a projection of QED in 3+1-dimensions to a 1+1-brane was performed. However, a finite-size regulator was introduced to avoid ultraviolet divergences that appear when confining the system to one dimension. Therefore, the effective theory obtained is not scale invariant and cannot be conformal. In our approach, we found an explicit way to deal with the divergences, such that we obtain the CQED without any regulator.
\end{thebibliography}
\end{document}